\documentclass{aa} 
\usepackage{graphicx}
\usepackage{txfonts}
\usepackage{float}
\usepackage{ulem}
\usepackage{hyperref}
\usepackage[title]{appendix}
\usepackage{academicons}
\usepackage{xcolor}
\usepackage{orcidlink}
\usepackage{url}
\usepackage{ulem}

\begin{document}

   \title{Dark Halos around Solar Active Regions. I. Emission properties of the Dark Halo around NOAA 12706. }

   \author{S.M. Lezzi\,\orcidlink{0000-0002-8506-1819} 
          \inst{1,2}, V. Andretta\,\orcidlink{0000-0003-1962-9741}\inst{1},
          M. Murabito\,\orcidlink{0000-0002-0144-2252}\inst{1}, \and
          G. Del Zanna\,\orcidlink{0000-0002-4125-0204}\inst{3}}

   \institute{Università di Napoli “Federico II”, C.U. Monte Sant’Angelo, Via Cinthia, I-80126 Naples, Italy\\
              \email{serena.lezzi@inaf.it}
         \and
             INAF - Osservatorio Astronomico di Capodimonte, Salita Moiariello 16, I-80131 Naples, Italy
        \and DAMTP, University of Cambridge, Wilberforce Rd, Cambridge CB3 0WA, UK
             }

   \date{Accepted 19 September 2023}


   \abstract
  {Dark areas around active regions (ARs) have been first observed in chromospheric lines more than a century ago and are now associated to the H$\alpha$ fibril vortex around ARs. Nowadays, large areas surrounding ARs with reduced emission relative to the Quiet Sun (QS) are also observed in spectral lines emitted in the transition-region (TR) and low-corona. For example, they are clearly seen in the SDO/AIA 171 {\AA} images. We name these chromospheric and TR/coronal dark regions as Dark Halos (DHs). Coronal DHs are poorly studied and, because their origin is still unknown, to date it is not clear if they are related to the chromospheric fibrillar ones. Furthermore, they are often mistaken for Coronal Holes (CHs).}
   {Our goal is to characterize the emission properties of a DH by combining, for the first time, chromospheric, TR and coronal observations in order to provide observational constraints for future studies on the origin of DHs. This study also aims at investigating the different properties of DHs and CHs and at providing a quick-look recipe to distinguish between them. }
   {We study the DH around AR NOAA 12706 and the southern CH, that were on the disk on 2018 April 22, by analyzing IRIS full-disk mosaics and SDO/AIA filtergrams to evaluate their average intensities normalized to the QS. In addition, we use the AIA images to derive the DH and CH emission measure (EM) and the IRIS \ion{Si}{IV} 1393.7 \r{A} line to estimate the non-thermal velocities of plasma in the TR. We also employ SDO/HMI magnetograms to study the average magnetic field strength inside the DH and the CH.}
  {Fibrils are observed all around the AR core in the chromospheric \ion{Mg}{II} h\&k IRIS mosaics, most clearly in the h$_3$ and k$_3$ features. The TR emission in the DH is much lower compared to QS area, unlike in the CH. Moreover, the DH is much more extended in the low-corona than in the chromospheric \ion{Mg}{II} h$_3$ and k$_3$ images.  Finally, the intensities, EM, spectral profile, non-thermal velocity and average magnetic field strength measurements clearly show that DHs and CHs exhibit different characteristics and therefore should be considered as distinct types of structures on the Sun. }
{}

   \titlerunning{Emission properties of the Dark Halo around NOAA 12706 }
   \authorrunning{Lezzi, S.M., Andretta, V., Murabito, M., \& Del Zanna, G.}
   \maketitle
%

\section{Introduction}
\begin{figure*}
    \centering
    \includegraphics[scale=1.,trim=83 360 30 47,clip]{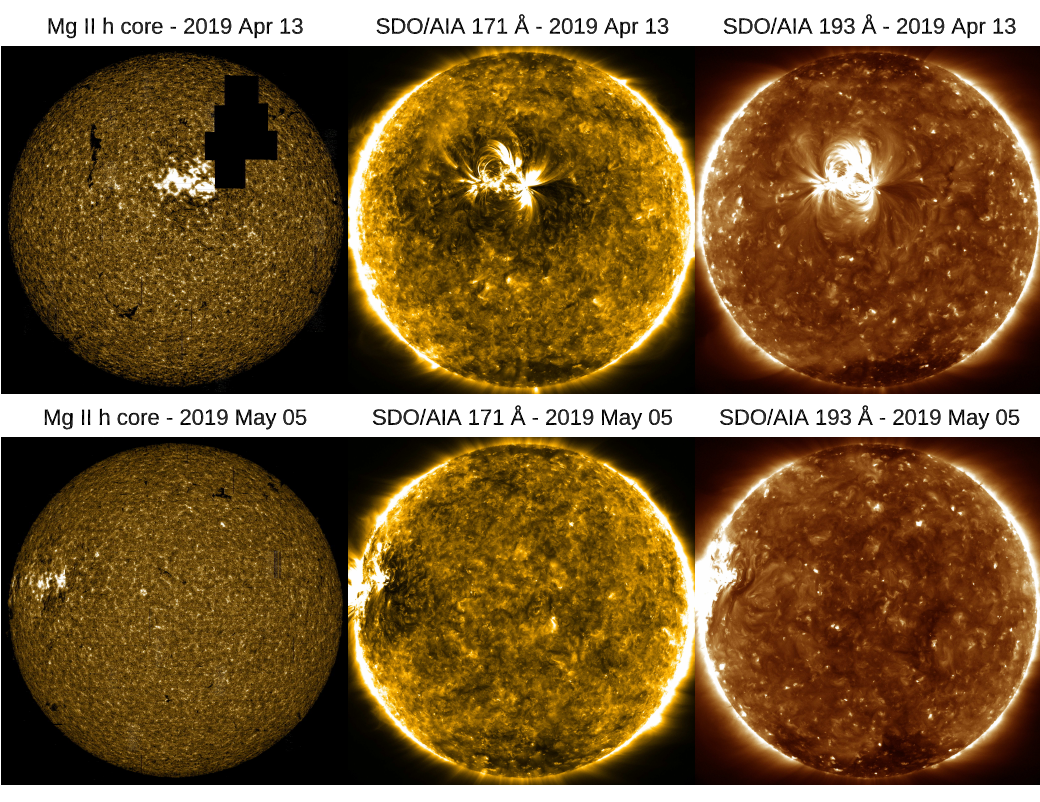}
    \caption{From left to right: IRIS Mg II h core full-disk mosaic, AIA 171 \r{A} and 193 \r{A} images for 13 April 2019 (top panels) and 05 May 2019 (bottom panels), when the ARs NOAA 12738 and 12740 respectively were on the disk. This two examples show the appearance of the DHs at the disk center and at the solar limb, respectively. DHs are dark areas best seen in TR/low-corona lines, such as \ion{Fe}{IX} in the AIA 171~\AA\ waveband. In the chromosphere, such as in \ion{Mg}{II} h\&k line cores, in correspondence of their inner part fibrils are often observed. }
    \label{fig1.0}
\end{figure*}
Active Regions (hereafter: ARs) on the Sun are typically surrounded by areas that in some wavelengths appear darker than the average Quiet Sun (hereafter: QS). They were first identified in the chromosphere as faint dark areas surrounding plages in the \ion{Ca}{II} K-line (\citealt{hale1903}; \citealt{stjohn1911}), and later also in the \ion{Ca}{II} 8542 \r{A} line \citep{dazumbuja1930}. \citet{deslanders1930} called them “circumfacules”.  \citet{bumba1965} found a spatial correlation between \ion{Ca}{II} K-line circumfacules and the H$\alpha$ fibril vortex around ARs and proposed that the \ion{Ca}{II} circumfacular regions are composed of dark H$\alpha$ fibrils. This suggestion was supported by many authors over the years (e.g. \citealt{veeder1970}; \citealt{foukal1971a}; \citealt{foukal1971b};
 \citealt{harvey2006}; \citealt{rutten2007};  \citealt{cauzzi2008}; \citealt{reardon2009}; \citealt{pietarila2009}).

Today, thanks to ultraviolet (UV) and extreme-ultraviolet (EUV) observations, it is possible to observe regions around ARs of reduced emission also in a wide range of spectral lines originating in Transition Region (hereafter: TR) and low corona.
For instance, areas of reduced emission around ARs can be seen in the Solar
Ultraviolet Measurements of Emitted Radiation (SUMER;
\citealt{wilhelm1995}) O V at 630 \r{A} and S VI at 933 \r{A} images\footnote{See, for instance, 07 June 1996 and 12 May 1996 images on SUMER atlas at \url{https://ads.harvard.edu/books/2003isua.book/}.}, as pointed out by \citet{andretta2014}. These authors refer to such structures as “dark halos” (hereafter: DHs) as they were extended dark areas as seen in full-Sun SOHO/CDS transition-region lines. We adopt this name in this work from now on. 

DHs are also clearly visible in the 171 \r{A} images taken by the Atmospheric Imaging Assembly (AIA; \citealt{lemen2012}) on board the Solar Dynamics Observatory (SDO; \citealt{pesnell2012}), as noted by \citet{wang2011}, who instead name them “dark canopies”. 
These latter authors studied the relationship between the 171~\r{A} DHs and the evolving photospheric field under the assumption that the 171 \r{A} DH is the counterpart of the H$\alpha$ fibrils, i.e. consists of EUV-absorbing chromospheric material (neutral hydrogen or helium) tracing out horizontal magnetic fields.
However, as already pointed out by \citet{andretta2014}, inspection of DHs using SUMER observations reveals that these regions are seen darker even in lines at wavelengths longer than the edge of the Lyman continuum at 912 \r{A}. Therefore, the interpretation based on absorption by neutral hydrogen cannot fully explain their low emission. \\
On the other hand, more recently, \citet{singh2021} argued that the existence of DHs, that they call instead “dark moats”, is related to the magnetic pressure coming from the strong magnetic field that splays out from the AR, pressing down and flattening to a low altitude the underlying magnetic loops. Those loops, which would normally emit the bulk of the 171 \r{A} emission, are restricted to heights above the surface that are too low to have 171 \r{A} emitting plasma sustained in them, according to \citet{antiochos1986}. According to this model, these flattened loops are expected to have temperatures less than $\sim$5 $\times$ 10$^4$ K. Therefore they should emit into the emission passband of the 304 \r{A} and appear bright in 304 \r{A} images. However, DHs are generally visible as darker areas in 304 Å images by visual inspection, which means that this model is not entirely consistent with observations. 
 \begin{figure*}
    \centering\includegraphics[scale=0.48,trim=15 10 15 5,clip]{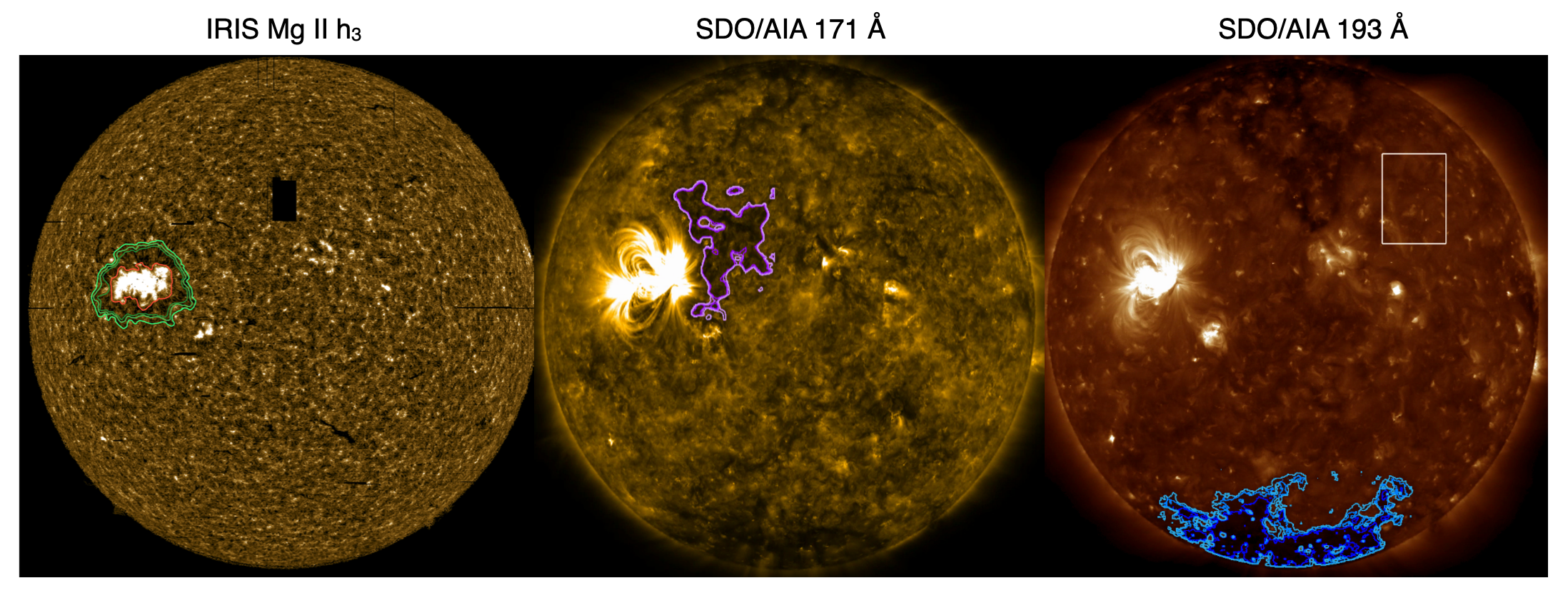}
    \caption{IRIS \ion{Mg}{II} h$_3$ full-disk mosaic and AIA full-disk images showing the DH surrounding AR NOAA 12706 in the northern hemisphere on 2018 April 22. The three images are center-to-limb corrected. Left panel: the three fibrillar DH contours are shown in shades of green and the outer AR-core edge is shown red. Middle and right panels: AIA 171 and 193 \r{A} filtergrams, respectively, taken at 19:24 UT. The three coronal DH contours are shown in shades of violet in the AIA 171 \r{A} image, while the QS box and the three southern CH contours are shown in white and in shades of blue, respectively, in the AIA 193 \r{A} image, as described in Section 3. The contours are obtained through intensity thresholds from the non center-to-limb corrected  images.}
    \label{fig1}
\end{figure*}
Consequently, at the moment the physical mechanism responsible for the DH reduced emission remains poorly understood. Moreover, it is unclear whether the coronal EUV-observed DH is related to the chromospheric fibrillar one, as there are no previous studies analyzing them with chromospheric, TR and coronal observations together. DHs are very common solar features visible around ARs independently from the solar cycle and at all distances from the disk center, as shown for instance in Fig. \ref{fig1.0}. 
DH could sometimes be mistaken for filament channels, which can be usually observed in the H$\alpha$ images (\citealt{makarov1982}). We have verified that for the ARs shown in Fig. \ref{fig1.0} and for AR NOAA 12706 that we have studied in this work no nearby filament channels are present in the H$\alpha$ images (see Fig. \ref{gong}), and therefore these DHs are not filaments channels.  
\\DHs have a spatial extent comparable to that of the associated ARs and therefore they may influence the irradiance of the Sun, especially during the maximum of the solar activity, when several ARs are present on the solar disk. 

Furthermore, DHs are sometimes mistaken for Coronal Holes (hereafter: CHs), and up to now there has not been a quantitative characterization that allows a clear distinction between these two solar features. \citet{andretta2014} noted that these halos are easily seen in TR lines, while CHs are not. \citet{singh2021} also observed that CHs tend to be dark across the six hotter AIA EUV wavelength passbands, in contrast with DHs, which are dark primarily in 171 \r{A}, but less so or not at all in other wavelengths, such as 193~\r{A}, as also shown in Fig. \ref{fig1.0}.

Here we use a combination of imaging, spectroscopic, and magnetic field observations to comprehensively characterize the DH around AR NOAA 12706. This multi-wavelength approach represents the first study of a DH through all the layers of the solar atmosphere, from the chromosphere up to the low corona. 
The aim of the work is to provide a quantitative characterization of the emission properties of a DH by using the Interface Region Imgaging Spectrograph (IRIS; \citealt{depontieu2014}) full-disk mosaics and SDO/AIA (\citealt{lemen2012}) images, in order to provide empirical constraints to the various hypotheses on the origin of DHs. In addition, our results show fundamental differences between the DH and the CH under study and support the thesis that DH and CH are two different types of structures on the Sun. Eventually, we provide a quick inspection method that can be used to distinguish a DH from a CH. 


\section{Observations}
We select AR NOAA 12706 for its suitable location on the disk in the IRIS mosaics: its intermediate position between the limb and the disk center makes it possible to look under the coronal loops when looking at AIA images. 
AR NOAA 12706 displays a $\beta$ magnetic configuration characterized by one sunspot of negative polarity and a more diffuse positive magnetic concentration as little dark pores. It was visible on the disk from 20 April to 28 April 2018. On April 20, a “light bridge” (see e.g. \citealt{vandriel2015}, \citealt{falco2016}, \citealt{felipe2016} and reference therein) begins to form within the negative polarity sunspot, indicating the onset of its fragmentation. Indeed, the spot completely vanishes on April 27. This AR was observed in a temporal interval of $\sim$ 4 hours by IRIS on 2018 April 22 during the full-disk observation mode, which started on April 22 10:41:50 UT and finished on April 23 04:05:45 UT. Southwest of NOAA 12706 is NOAA 12707 (see e.g. Figs. \ref{fig1} and \ref{fig2c}), a smaller decaying active region that disappears from the disk in about two days. Given its small size, we assume that its presence does not influence the area around AR 12706 and therefore we do not include it in the analysis. The same IRIS \ion{Mg}{II} mosaics have been studied previously by \citet{bryans2020} and the AR NOAA 12706 is one of the seven ARs studied by \citet{singh2021}. We will discuss our results in relation to those of these authors in Sect. 5.

\subsection{IRIS full-disk mosaics}
Primarily used for quasi-regular monitoring of changes in instrument sensitivity, IRIS full-disk mosaics have proven to be very useful in the study of the solar atmosphere, as they have been employed in a variety of ways (see e.g. \citealt{schmit2015}; \citealt{vial2019}; \citealt{bryans2020}; \citealt{ayres2021}; \citealt{koza2022}).
IRIS full-disk mosaics consist of spectral maps of the entire solar disk built through an observing sequence (currently run approximately once per month, when IRIS is not in eclipse season) that takes a series of successive rasters at 184 different pointing locations. Rasters' spectral windows include the six strongest spectral lines, i.e. the resonance lines \ion{Mg}{II} h and k ($T\sim8 \times 10^3$ K), \ion{C}{II} 1335 and 1334 \r{A} ($\sim 10^4$ K), and \ion{Si}{IV} 1393 and 1403 \r{A} ($\sim6 \times 10^4$ K). All 184 positions take $\sim 18$ hr, including $\sim 10$ hr of time
taken to repoint the spacecraft (\citealt{schmit2015}). Each individual observation consists of a 64-step raster with 2$''$ steps and 2 s exposure time at each slit position. The spectra along the slit have been binned to a resolution of 0.33$''$. \ion{Mg}{II} mosaics have a spectral scale of 35 m\r{A} px$^{-1}$, while \ion{C}{II} and \ion{Si}{IV} have a spectral scale of 25 m\r{A} px$^{-1}$. Thus the spectral range of 3.5 \r{A} for \ion{Mg}{II} lines and of 1 \r{A} for \ion{C}{II} and \ion{Si}{IV} lines is sampled by 101 and 41 spectral points, respectively.

For this work, mosaics represent a unique resource to investigate the morphological features of DHs, which have a spatial extent comparable to that of the associated ARs and could not be properly studied through normal rasters, given their usually relatively small FoV. In addition, the available spectral windows allow a view in both chromosphere and TR, thanks to the chromospheric \ion{Mg}{II} h\&k lines and the \ion{C}{II} and \ion{Si}{IV} TR doublets.

\subsection{SDO data}
In order to characterize the emission properties of the DH in the corona, we use AIA observations for their full-Sun coverage. DHs are relatively stable structures that evolve in timescales of the order of the AR evolution time ($\sim$ days, see e.g. \citealt{vandriel2015}). Therefore, rather than analysing all the AIA images in the four-hour period taken by IRIS to entirely cover the AR by rasters, we study only the AIA image taken at 19:24 UT, which is approximately at the middle of the $\sim$ 4 hr temporal window.

To study the evolution of the magnetic flux along the line of sight, we also use a series of Helioseismic and Magnetic Imager (HMI; \citealt{schou2012}) LOS magnetograms taken in the Fe I 617.3 nm line with a resolution of 1$"$. They cover $\sim$ 30 mins of
observations, starting from the 19:07:30 UT to 19:36:45 UT with a cadence of 45 s.

\begin{figure*}
\centering
    \includegraphics[scale=0.75,trim=13 50 28 195, clip]{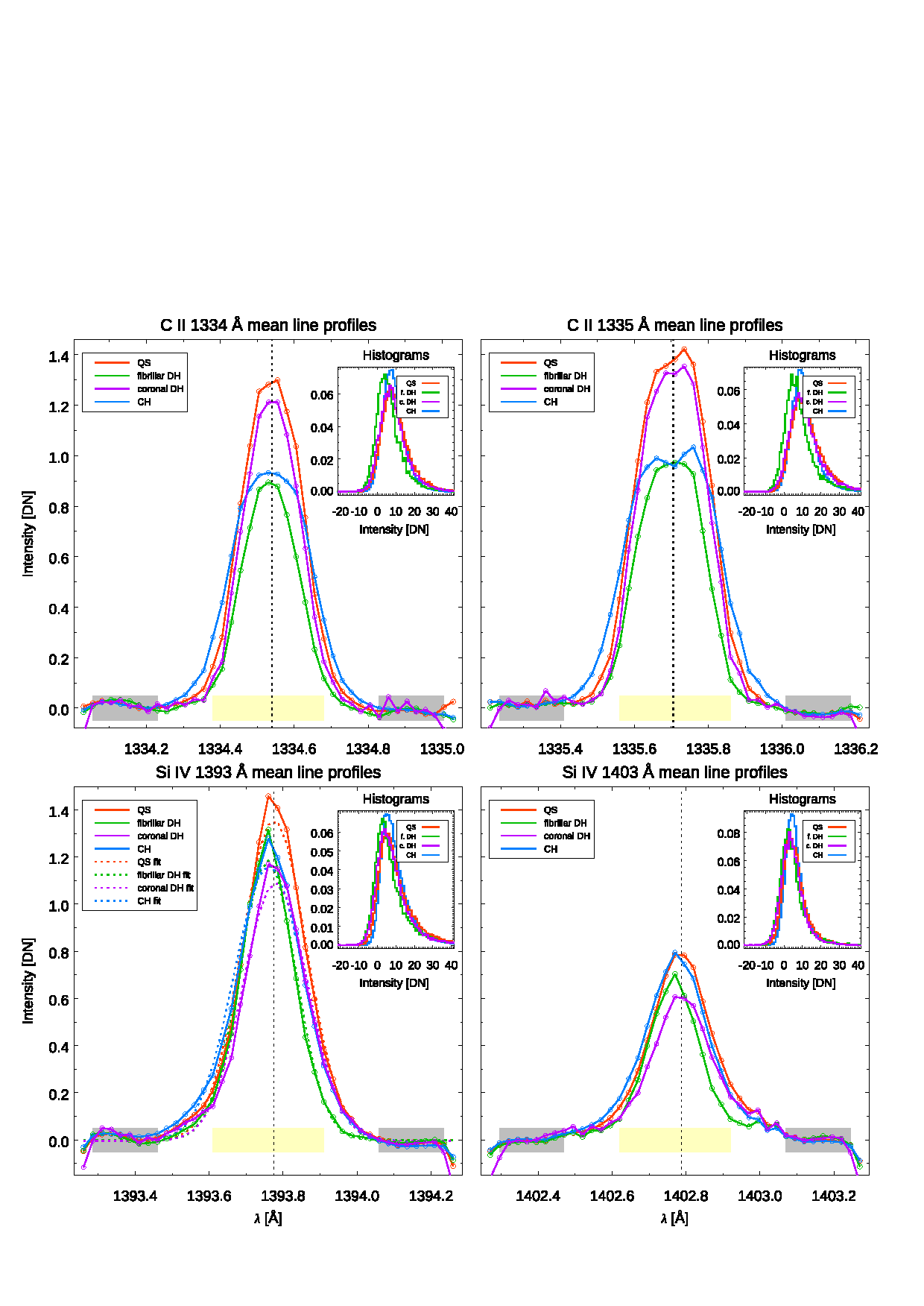}
    \caption{\ion{C}{II} 1334 \& 1335 \r{A} (upper panels) and \ion{Si}{IV} 1393 \& 1403 \r{A} (lower panels) average profiles for QS, fibrillar and coronal DH and CH, in red, green, violet and cyan respectively, with the corresponding normalized histograms of the integrated intensities. The vertical dotted lines in the profiles represent the QS reference wavelengths. The yellow and gray bands identify the spectral intervals used for \ion{C}{II} and \ion{Si}{IV} intensity and background evaluation, respectively. In the lower left panels, the dashed curves are the \ion{Si}{IV} 1393 \r{A} fits.
    }
    \label{figlines}
\end{figure*}

\begin{figure}
    \centering
    \includegraphics[scale=0.96,trim=5 140 380 250, clip]{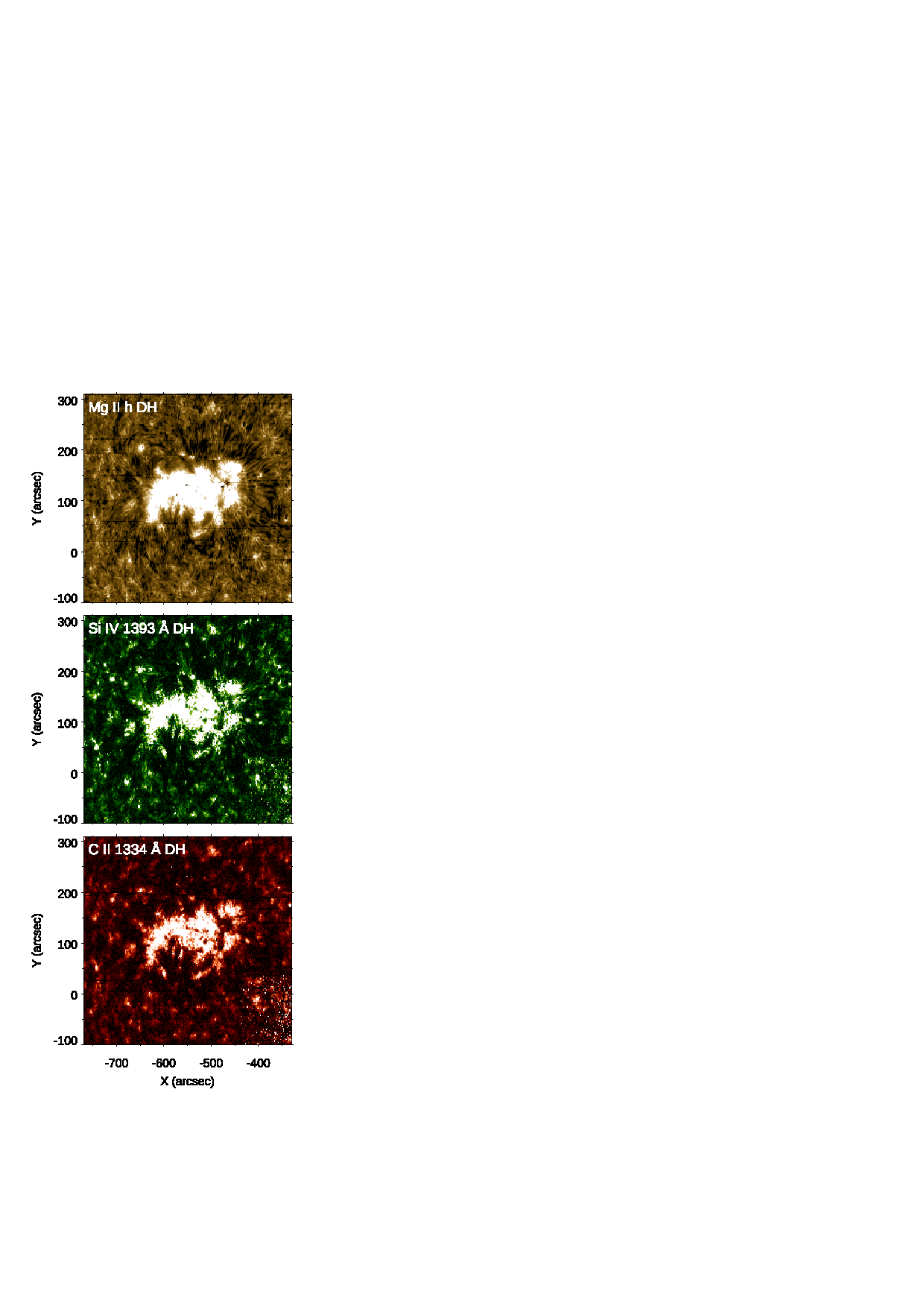}
    \caption{Close-up view of NOAA 12706 and its surroundings. From top to bottom: IRIS \ion{Mg}{II} h$_3$, \ion{Si}{IV} 1393 \r{A} and \ion{C}{II} 1334 \r{A}. North is up and west is to the right. FoV has dimensions 440" $\times$ 410". In the \ion{Mg}{II} h$_3$ it is possible to observe around the AR core the chromospheric DH made up of fibrils, that have maximum length of approximately the AR half size in the longitudinal direction. In the \ion{C}{II} and \ion{Si}{IV} mosaics, instead, it is still recognizable a halo of reduced emission around the AR, but less clearly and without the evident presence of fibrillar structures. 
}
\label{fig2a}
\end{figure}

\begin{figure}
    \centering
    \includegraphics[scale=0.94,trim=318 140 0 170, clip]{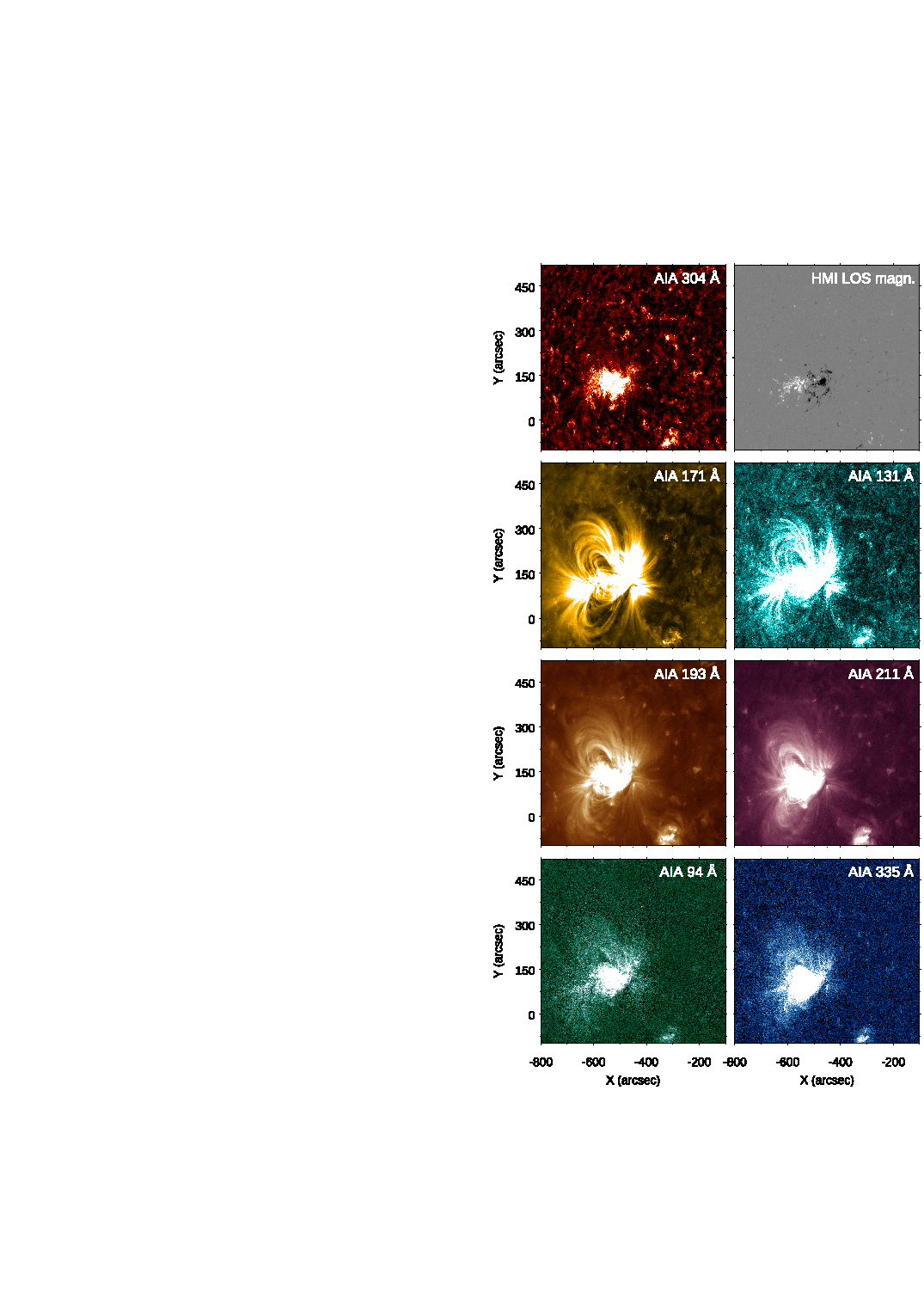}
    \caption{AIA close-up views of NOAA 12706 and its surroundings. North is up and west is to the right. FoV has dimensions 700" $\times$ 620".}
\label{fig2c}
\end{figure}

\section{Analysis}
To quantitatively characterize the emission properties of the DH in the different layers of the solar atmosphere and compare them to the QS and the CH properties, we define four regions of interest (hereafter: RoIs), described below.
For the DH, the starting point of the analysis is the annular region whose emission is reduced compared to the QS  seen by eye in the IRIS \ion{Mg}{II} h\&k core full-disk mosaics (e.g. see left panel of Fig. \ref{fig1} and in top panel of Fig. \ref{fig2a}). In particular, fibrils are clearly observed in the \ion{Mg}{II} h\&k cores (i.e. 50$^{th}$ spectral point for both lines, see right panel of Fig. \ref{fibrils}). Fibrils are chromospheric features visible mainly in nearline-core wavelengths of the chromospheric lines (see e.g. \citealt{kianfar2020}). Indeed, also in IRIS \ion{Mg}{II} mosaics it is possible to notice that they become less visible as the selected wavelength approaches the line wings, where they become fainter and eventually undetectable (see Fig. \ref{fibrils}), meaning that they are upper chromospheric features. 
\\Because the fibrils around the AR are most visible at the \ion{Mg}{II} h\&k cores, we use in our analysis only the h$_3$ and k$_3$ images\footnote{We use the well-known nomenclature to define $h_3$ as the central depression of the line, $h_{2r}$ as the red emission peak and the $h_{1r}$ as the minimum on the red side of the central emission peak. These positions are shown as example in the \ion{Mg}{II} h full-disk mean profile of Fig. \ref{fibrils} (right panel).}, found through a double gaussian fit of the \ion{Mg}{II} h\&k  lines.
 We use the \texttt{defroi.pro} IDL routine for manually drawing, in the \ion{Mg}{II} h$_3$ mosaic, the outer boundary of the DH, visually following the fibrils, and the outer edge of the AR core that is the inner DH contour. These two contours are shown in the left panel of Fig. \ref{fig1} in green and red respectively. We will refer to the DH identified by the these two contours as “chromospheric fibrillar DH”.
We point out that the close NOAA 12707, on the contrary, does not exhibit a chromospheric DH made up of fibrils extending radially from the AR core, but it shows few dark filamentous structures  surrounding it.

Looking at AIA filtergrams (see Fig. \ref{fig2c}), we notice that the DH is most visible in the AIA 171 \r{A} image, as also shown by \citet{wang2011} and \citet{singh2021}. In this waveband, the DH spatial extent is larger than in the \ion{Mg}{II} mosaics, with coronal loops immediately close to the AR core partially covering and hiding the dark emission beneath them. Therefore, we decide to use a second contour for the coronal 171 \r{A} DH, which we call “coronal DH”, located on the west side of the AR. We define it through an intensity threshold equal to $55\%$ of the 171 \r{A} average disk emission. We use this particular value because the area falling inside the corresponding contour is close to the dark region identifiable by eye. This new contour, used to evaluate the coronal DH average intensity, is shown in violet in middle panel of Fig. \ref{fig1}.  We point out that in this image projection effects prevents the visibility of the east side of the coronal DH, which is instead well observed following the AR during its transit on the solar disk. 

For the CH definition, we use another intensity threshold, equal to the $35\%$ of the 193 \r{A} average disk emission. Indeed, tools based on intensity threshold of coronal EUV images have been largely used for the detection of CHs in the past (e.g. \citealt{Krista2009,Reiss2015,Heinemann2019}). The CH contour is shown in blue in Fig. \ref{fig1}. 
For the QS we use a box of $250" \times 337"$ located outside CH and AR areas (see right panel in Fig. \ref{fig1}) and chosen such that its radial distance from the center is approximately similar to that of the AR, in order to minimize center-to-limb variations. 
In summary, we have defined four RoIs, shown in Fig. \ref{fig1}:
\begin{itemize}
    \item chromospheric fibrillar DH;
    \item upper coronal DH;
    \item southern CH;
    \item QS.
\end{itemize}

\subsection{Data processing}
The AIA images have been processed by first deconvolving the psf with \texttt{aia\_deconvolve\_richardsonlucy.pro} IDL routine.
All IRIS mosaics are first of all processed using the \texttt{new\_spike.pro} IDL routine in order to remove spikes. In the \ion{C}{II} and \ion{Si}{IV} mosaics an anomalous background is present and it is thought to be a calibration issue, likely related to incomplete compensation for stray light contamination (see \citealt{ayres2021} for further details). For this reason a background has been calculated and subtracted, as also done by \citet{ayres2021}. In particular, for every spatial pixel, we evaluate the background $I_{bkg}$ as follow:
\begin{equation}
\centering
   I_{bkg}= 0.5\cdot\Biggl(\sum_{i_{red}=1}^{8} I_{i_{red}} + \sum_{i_{blue}=1}^{8} I_{i_{blue}}\Biggr)
    \label{eqlines}
\end{equation}
where $i_{red}\in [\lambda_{ref}+12\Delta\lambda, \lambda_{ref}+19\Delta\lambda]$, $i_{blue}\in [\lambda_{ref}-19\Delta\lambda, \lambda_{ref}-12\Delta\lambda]$, and where $\lambda_{ref}$ represents the QS reference wavelength and $\Delta\lambda=0.025$ \r{A} corresponds to the spectral scale of \ion{C}{II} and \ion{Si}{IV}. 
The \ion{C}{II} and \ion{Si}{IV} background spectral intervals are shown as gray bands in Fig. \ref{figlines}.

Then we correct the line intensities for the center-to-limb variation, as described in Appendix. In Fig. \ref{figlines} we show the center-to-limb corrected average line profiles for \ion{C}{II} and \ion{Si}{IV} doublets inside the four RoIs, i.e. QS, fibrillar DH, coronal DH and CH, with the corresponding normalized histograms of the integrated intensities.

For \ion{C}{II} and \ion{Si}{IV}, we use the mosaics after the above processing  computed integrating the lines on the spectral intervals shown as yellow bands in Fig. \ref{figlines}. 
In Fig. \ref{fig2a} we show a close-up view of AR NOAA 12706 as observed, from top to bottom, in \ion{Mg}{II} h$_3$ and in the background-removed and peak-integrated \ion{Si}{IV} 1393 \r{A} and \ion{C}{II} 1334 \r{A} center-to-limb corrected mosaics. These zooms are representative also for the \ion{Mg}{II} k$_3$, \ion{Si}{IV} 1403 \r{A} and \ion{C}{II} 1335 \r{A}, respectively, that we do not show because they appear remarkably similar.

\subsection{Average intensity measurements}
In order to quantitatively characterize the emission properties of the fibrillar and the coronal DHs, we evaluate the average intensity (in DN) inside the four RoIs. We apply the contours defined before as follows:
\begin{itemize}
    \item the annular \ion{Mg}{II} h$_3$ fibrillar DH contour is applied to all IRIS mosaics;
    \item the AIA 171 \r{A} coronal DH contour is applied to all AIA filtergrams;
    \item the same AIA 193 \r{A} CH contour is applied to all IRIS mosaics and AIA filtergrams;
    \item the same QS box is applied to all IRIS mosaics and AIA filtergrams.
\end{itemize}
Then we normalize the fibrillar DH, the coronal DH and the CH average intensities to the QS, by computing the ratio defined as:
\begin{equation}
    \mathrm{ratio} = \dfrac{I_{av,RoI}}{I_{av,QS}}, 
    \label{ratioeq}
\end{equation}
where the RoIs are the fibrillar DH, the coronal DH and the CH. 

The uncertainties on the ratios defined in the above Eq.~\ref{ratioeq} are in principle the result of the propagation of the uncertainties attached to each step of the analysis, i.e. the definition of the RoI, the calculation of the average intensity within the RoI, and the center-to-limb correction.  In the case of intensities from IRIS spectra, we considered in addition the uncertainty due to the definition of the spectral intervals used for evaluating the background and the total intensity, and the contribution due to the noise in each spectral pixel of the profile. We evaluated each one of these sources of uncertainties, and found that by far the largest contribution is due to the RoI selection criteria.

In order to estimate the uncertainties on the ratio given by Eq.~\ref{ratioeq}, we then proceed as follows:
\begin{itemize}
    \item For the fibrillar DH we drew by hand two additionally contours by being more and less restricted on the fibril-edge definition. These contours are shown in different shades of green in the left panel of Fig. \ref{fig1}.
    \item For the coronal DH and CH, we changed the value of the intensity thresholds defining the RoIs; in particular, we considered a 2\% variation in the thresholds defining the coronal DH RoI, i.e. we considered two additional contours at 55$\%$ and 59$\%$ of the mean disk value, and a 10\% variation for the CH RoI, i.e. at 25$\%$ and 45$\%$ of the mean disk value. We thus obtained for each of those RoI two additional regions of slightly smaller and larger areas, respectively, than the reference RoI. The resulting three coronal DH contours are shown in violet in the middle panel of Fig. \ref{fig1}; the three CH contours are shown in blue in the right panel of Fig. \ref{fig1}.
\end{itemize}
The uncertainties of the ratios of Eq.~\ref{ratioeq} are then obtained by computing the difference of the mean intensities in these additional regions with respect to the value obtained in the reference RoI.

\begin{figure*}
\centering 
\includegraphics[scale=0.5,trim=130 0 0 25, clip]{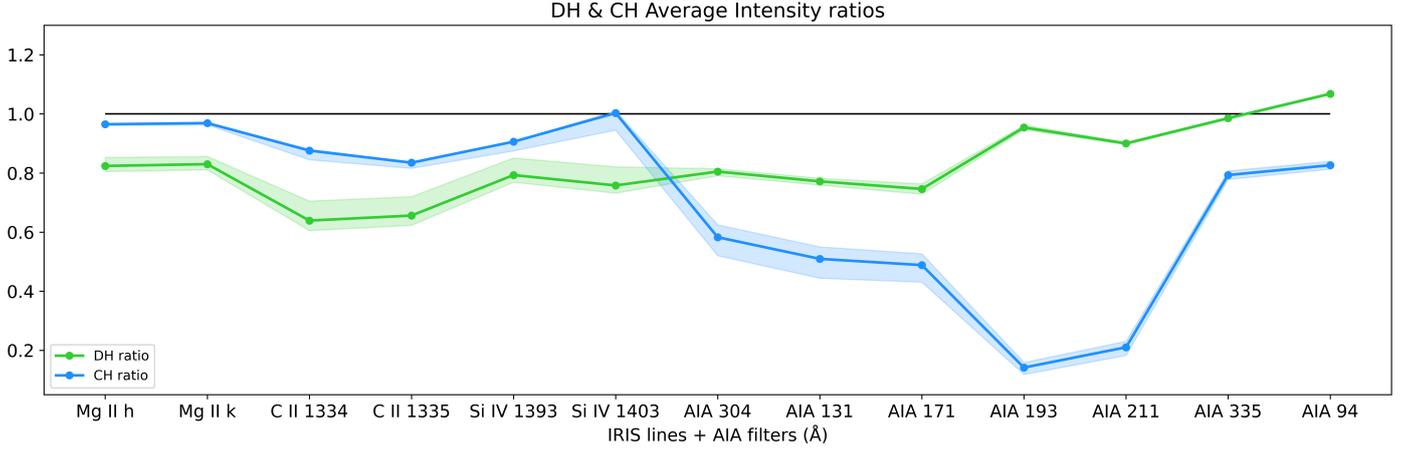}
    \caption{DH (in green) and southern CH (in blue) average intensity ratios in IRIS lines and AIA filters. The shaded areas are the ratio uncertainties, computed as discussed in Section 3. The chromospheric fibrillar and the 171 \r{A} coronal DH contours have been employed for the IRIS and AIA ratios, respectively. }
    \label{figratios}
\end{figure*}

The ratios for the selected RoIs and the associated uncertainties computed as described above, are shown in Fig. \ref{figratios}. The dispersion of the ratios shown with shaded green and blue bands is obtained using the additional contours of slightly different extent to give an estimate of the ratio variability.  These results will be discussed in more details in Sec.~\ref{sec:results}.

\begin{table}[]
    \centering
    \begin{tabular}{c c c c c c }\hline \hline 
              &    $I_0$ [DN]   & $\lambda_0$ [\r{A}] & $\sigma$ [\r{A}] & $I_1$ [DN] & $\xi$ [km/s]\\\hline
        f.DH   &    1.18  &   1393.76   &  0.077  &  -0.0001 &  22.2$\pm$0.2\\
        c.DH   &    1.09    &   1393.78   &  0.088  &  -0.003 &  25.7$\pm$0.5\\
        CH   &    1.17    &   1393.76   &  0.094  &  -0.004 &  27.8$\pm$0.7\\
        QS   &   1.34    &   1393.77   &  0.087  &  -0.00001 & 25.6$\pm$0.4\\
        \hline\hline\\
    \end{tabular}
    \caption{\ion{Si}{IV} 1393 \r{A} gaussian fit parameters inside the fibrillar DH, the coronal DH, the CH and the QS; $\xi$ is the non-thermal velocity retrieved using Eq. \ref{eqsigma}. 
    }
    \label{ntvel_table}
\end{table}

\subsection{Non-thermal velocity measurements}
In the TR the profiles of the spectral lines often exhibits, in addition to the thermal and instrumental broadenings, excess broadening which is associated to unresolved plasma motions that can transfer mass and energy between the chromosphere and the corona. This plasma flows are likely to be intimately related to the still unknown physical processes occurring in the TR. We study the non-thermal broadenings inside the four defined RoIs using the optically thin line \ion{Si}{IV} 1393.7 \r{A}.
For optically thin lines the spectral profile is assumed Gaussian and the Gaussian width $\sigma$ can be expressed as (\citealt{delzanna2018}):
\begin{equation}
    \sigma^2=\dfrac{\lambda_0^2}{2c^2}\Biggl( \dfrac{2kT_i}{M}+\xi^2\Biggr)+\sigma_I^2,\label{eqsigma}
\end{equation}
where $c,k,T_i,\sigma_I,\xi$ are the speed of light, the Boltzmann constant, the temperature of the ion, the additional instrumental broadening, which is 3.9 km s$^{-1}$ for IRIS (\citealt{delzanna2018}, \citealt{depontieu2014}), and the non thermal velocity, which represents the most probable velocity of the random plasma motions. 

We estimate $\sigma$ inside the QS, the fibrillar DH, the coronal DH and the CH by fitting the \ion{Si}{IV} 1393 \r{A} mean line profiles shown in Fig. \ref{figlines} with a gaussian and a costant term. 
We assume that the \ion{Si}{IV} formation temperature of $60 \times 10^3$ K. This is lower than the temperature of $80 \times 10^3$ K obtained in CHIANTI using the zero-density ionization equilibrium (\citealt{dere2019}), and is derived with the new ionization equilibrium calculations by \citealt{dufresne2021} that include density-dependent effects and charge transfer.
The fits for the four RoIs are shown in bottom left panel of Fig. \ref{figlines}. The fit parameters and non-thermal velocities $\xi$ found are reported in Table \ref{ntvel_table}. The fit parameters $I_0$, $\lambda_0$, $I_1$ and $\sigma$, are the height, the center, the constant term and the standard deviation of the Gaussian, respectively.

\subsection{Emission measure}
To deduce the temperature distribution of the coronal plasma along the LOS from the AIA filtergrams, we calculate inside the coronal DH, the QS and the CH the average total column emission measure function $EM$, defined as 
\begin{equation}
    EM=\int_h N_e^2 dh= \int_T DEM(T) dT,\quad\quad\quad\quad\quad\quad\quad [cm^{-5}]
    \label{em}
\end{equation}
where $N_e$ is the electron number density and $DEM(T)$ is the column differential emission measure, defined as $DEM(T) = N_e^2 (dT/dh)^{-1}$. 

We calculate the $DEM(T)$ function employing the sparse inversion technique by \citet{cheung2015}, which assumes that the intensity in each pixel in an image from the $j$-th AIA waveband can be written as
\begin{equation}
    I_j=\int_0^{\infty} K_j(T)DEM(T)dT,
\end{equation}
where $K_j(T)$ is the function which describes the response of the $j$-th channel with the temperature.
This method inverts the previous equation obtaining $DEM$ solutions after giving as input the EUV images from the AIA channels 94, 131, 171, 193, 211, and 335 \r{A} and
the temperature response functions of these channels. We implemented this technique through the routines \texttt{aia\_sparse\_em\_init.pro} and \texttt{aia\_sparse\_em\_solve.pro} available in IDL SolarSoft.  The first one builds up the response functions, while the second one is the proper $DEM$ solver. 

Then we obtained the average emission measures using Eq. \ref{em} by integrating the $DEM$ over the temperature range from $log$T[K] = 5.6 to $log$T[K] = 6.3. We do not evaluate the average emission measure inside the fibrillar DH because AR loops cover that RoI.

\begin{figure}
\centering
\includegraphics[scale=0.9,trim=22 54 320 525, clip]{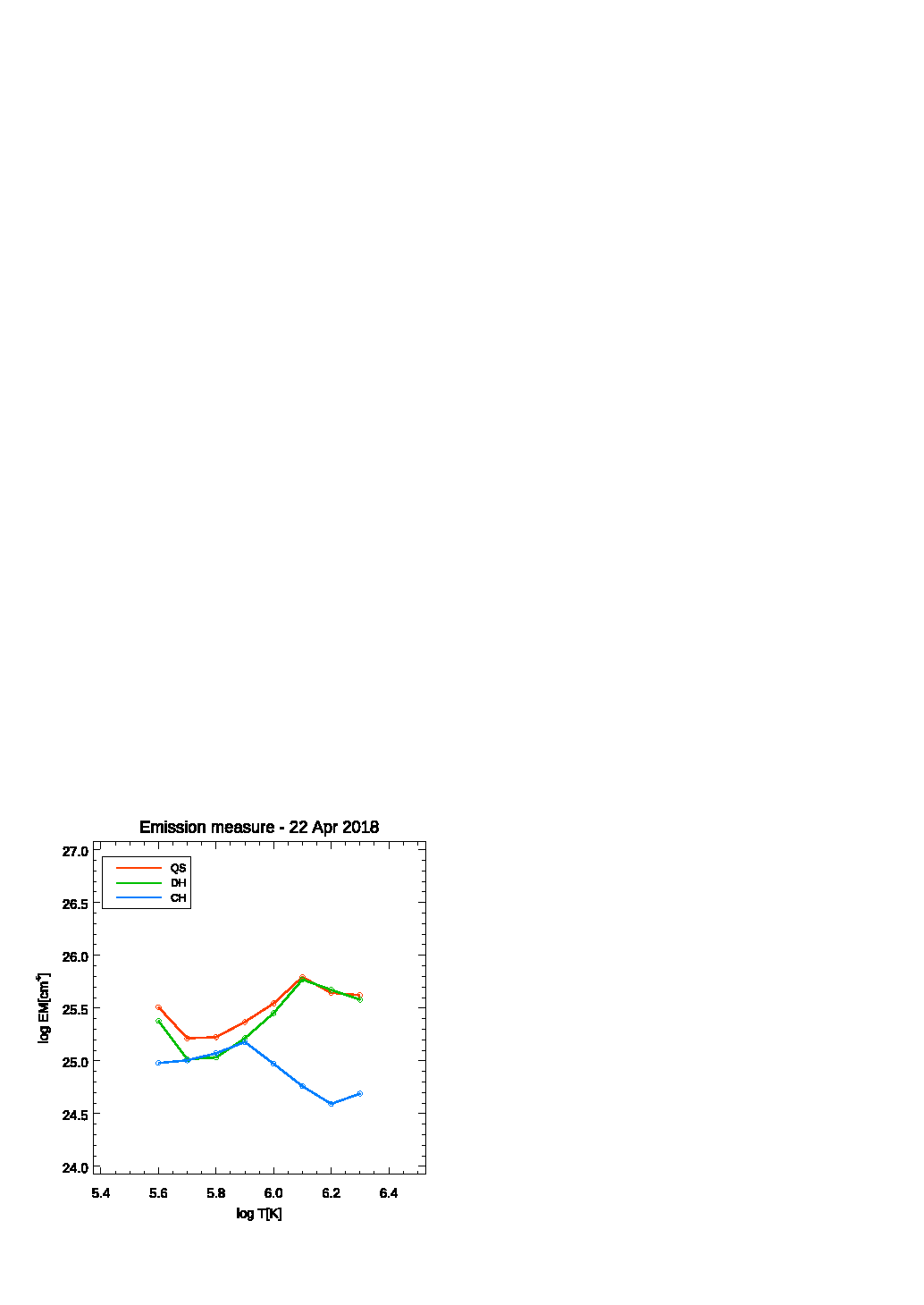}
    \caption{Average log EM curves as functions of log T calculated over the QS, coronal DH and southern CH. The DH and the CH show a similar depletion in the EM compared to the QS in the 5.6-5.9 log T range. At higher temperatures the CH exhibits a trend opposite to that of DH and QS, with a strong EM reduction. }
    \label{fig_em}
\end{figure}

\subsection{HMI magnetic field strengths}
\begin{figure}
\centering 
\includegraphics[scale=0.73]{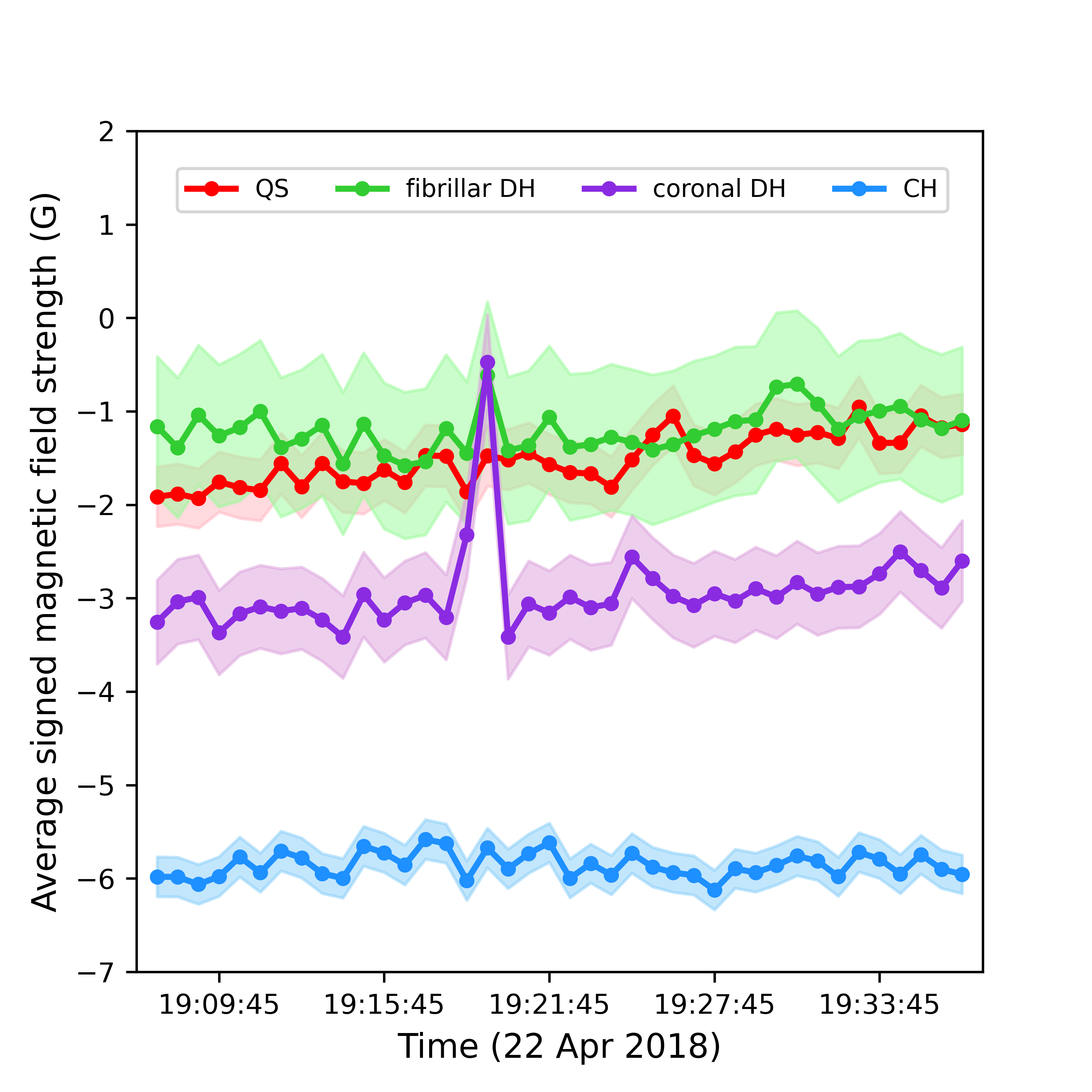}
\includegraphics[scale=0.73]{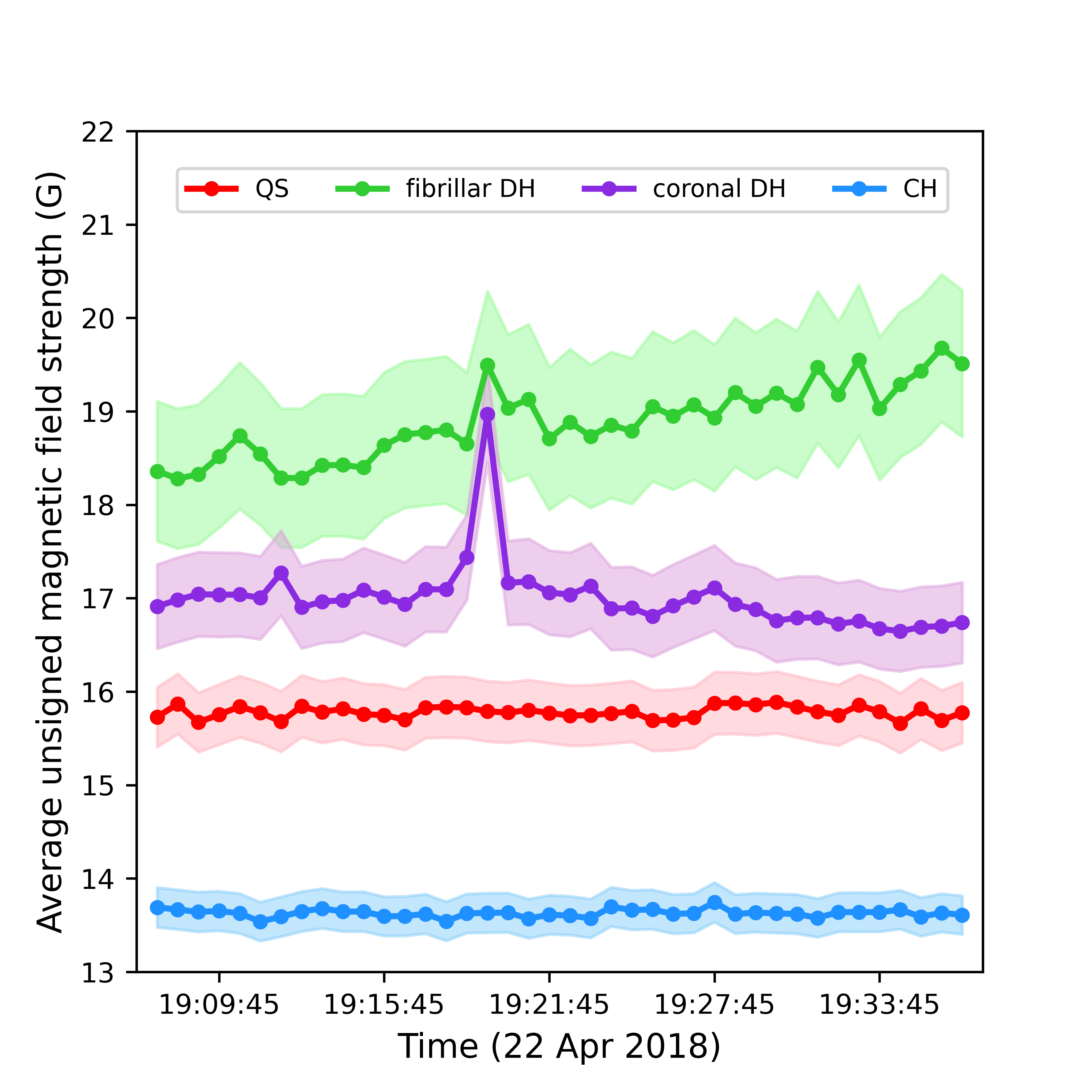}
    \caption{Average signed (top panel) and unsigned (bottom panel) magnetic field strength $\sim$ 30 minutes time evolution inside the four RoIs. The colored bands are the uncertainties.}
    \label{fig_hmi}
\end{figure}

We use the time series of HMI magnetograms to investigate the temporal evolution of the average signed $<B_{LOS}>$ and unsigned $<|B_{LOS}|>$ magnetic field strength inside the four RoIs. The noise level
of 45-s magnetograms exhibits a center-to-limb variation, being lower near disk center and increasing radially approaching the limb up to $\sim$ 8 G (\citealt{yeo2013}). Therefore, we cut all the pixels in the range [-8,8] G. The results of the average  signed and unsigned magnetic field strength measurements, together with standard deviations, are reported in Fig. \ref{fig_hmi}.  We point out that the magnetic field strength depends on the threshold used in the noise removal. However, we found that the relative distances between the RoIs remain the same also when using different thresholds.

\begin{figure}
    \centering
    \includegraphics[scale=0.7,trim=170 50 400 500]{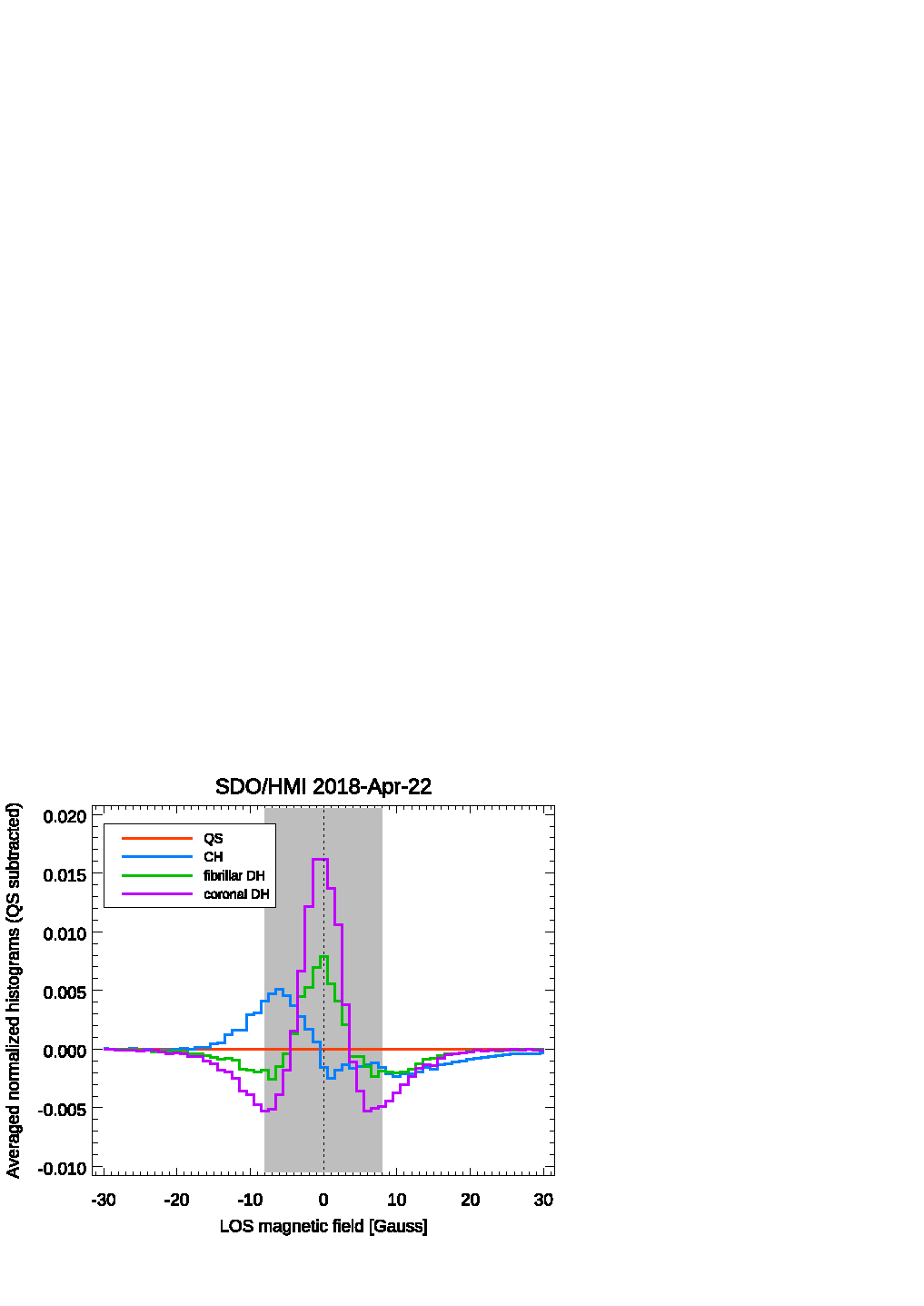}
    \caption{Averaged QS-subtracted histograms of the four RoIs. The gray band corresponds to the noise interval [-8,8] G that has been cut to obtain the magnetic field strength measurements shown in Fig. \ref{fig_hmi}.}
    \label{hmi_histo}
\end{figure}

\section{Results}\label{sec:results}
The first results we show are the \ion{C}{II} 1334 \& 1335 \r{A} (upper panels) and \ion{Si}{IV} 1393 \& 1403 \r{A} (bottom panels) mean line profiles of Fig. \ref{figlines}. The QS, fibrillar DH, coronal DH and CH profiles are shown in red, green, violet and cyan, respectively, after the background removal and the center-to-limb correction.
The gaussian \ion{Si}{IV} fits are added in the bottom panels of Fig. \ref{figlines}. At the same time, we show the histograms of the integrated intensity (in DN) inside the RoIs, normalized to the number of spatial pixels. We do not show the \ion{Mg}{II} profiles because in the analysis we use only the h$_3$ and k$_3$ images. 

The \ion{Si}{IV} lines have Gaussian profiles, while the \ion{C}{II} lines are flatter than a Gaussian, as also reported in \citet{upendran2021} for the CH and QS cases. We also notice a hinted double peak in the \ion{C}{II} CH line. In all cases, the DH presents narrower lines and intensity peaks slightly higher than the CH (except for the \ion{Si}{IV} 1403 \r{A} line). In the four lines the QS is more intense than the other RoIs, with the only exception of the coronal DH, that is nearly identical to the QS in the \ion{C}{II} lines. Overall, we do not find evident Doppler shifts in the profiles.

The histogram shapes are unimodal and have a broader right side, probably due to bright points. The four lines display a similar histogram of intensities scheme: the DH distribution (green) is shifted towards smaller intensity counts and has a peak value between those of the QS and the CH.

Regarding the non-thermal velocities obtained with the gaussian \ion{Si}{IV} 1393 \r{A} fit, the results of which are shown in Table \ref{ntvel_table}, we find that the plasma has non-thermal velocities  $\xi$ of $\sim$ 22, 26, 26 and 28 km/s  inside the fibrillar DH, coronal DH, QS and CH, respectively, with the coronal DH having essentially the same $\xi$ of the QS. The QS value is close to the averaged values found by \cite{rao2022}. 

In Fig. \ref{figratios} we show the DH (in green) and CH (in blue) ratios for the IRIS lines and the AIA bands. The IRIS lines are sorted according to the nominal formation temperature as determined from the standard CHIANTI ionization equilibrium.  The IRIS and AIA DH ratios are referred to the fibrillar and the coronal DH, respectively.
A fibrillar DH is observed in all IRIS lines, because IRIS ratios are less than one, and it is darkest in the \ion{C}{II} lines. The CH ratios display a similar trend but with higher ratios, closer to 1. In the AIA side of the plot, the coronal DH shows ratios less than one in all filters, except for the AIA 94 \r{A}, with a positive trend increasing with temperature. On the other side, the CH shows evidence of different behaviour: the CH has a drop in the intensity relative to the QS, which reaches a minimum in the AIA 193 and 211 \r{A} bands and then increase in the hottest AIA bands. Therefore, the fibrillar  and coronal DHs show approximately constant ratios of the order of 0.8, which increase approaching the hottest AIA bands. On the contrary, the CH exhibits an opposite behaviour, being almost not visible in the chromosphere and TR and becoming evident approaching the corona.

Fig. \ref{fig_em} shows the average EM as function of temperature obtained using the AIA filtergrams for the QS, the coronal DH and the CH in red, green and cyan respectively. The plot shows clearly that in the temperature range $log$ T[K] $=5.6-6.1$ both the coronal DH and the CH have a lower emission measure compared to the QS, while in the range $log$ T[K] $=6.1-6.3$, corresponding to the AIA channels 193 and 211 \r{A}, the DH and the CH have again opposite behaviour. We note that the DH and QS emission measure results in the temperature range $log$ T[K] $=5.6-6.3$ are comparable with those found by \cite{singh2021}, even if the RoIs are selected differently and the center-to-limb effect is not corrected.

In Fig. \ref{fig_hmi} we plot the average signed and unsigned magnetic field strength in the four RoIs obtained from the $\sim$ 30-minute sequence of magnetograms. We find that during the selected time interval the magnetic field strengths remain constant, and only the fibrillar DH displays a weak time evolution, probably related to new flux emerging in the RoI. We find that the fibrillar DH and the QS have similar averaged signed magnetic fields strengths, close to zero (consistent with the results by \cite{wang2011}), while the CH is unipolar, and the coronal DH presents values intermediate between the CH and the QS. On the other side, the fibrillar and coronal DH show average unsigned magnetic field strengths higher than the QS, and the CH displays the lowest values. 

In Fig. \ref{hmi_histo} we also show the 30-minute averaged QS-subtracted normalized histograms, which have been obtained by first normalizing the 30-minute averaged histograms to the total number of pixels and then by subtracting the QS histogram. We find that the fibrillar and coronal DH histograms are centered on the zero, while the CH histogram peak is shifted towards negative magnetic field strength values.

\section{Discussion and conclusions}
{

Dark halos (DHs) have for a long time been observed as ``circumfacular regions'' surrounding active regions, but have also been known under different names.  Typically, they appear as regions darker than the quiet Sun in chromospheric or TR lines.  These large-scale structures are normally best studied in full-Sun imaging or imaging-spectroscopy.   We thus inspected the available set of IRIS full-Sun mosaics (2013 to present), and found that most ARs surroundings exhibit clear DH signatures in the core of the \ion{Mg}{II} h \& k lines, appearing as dark fibrils with almost radial symmetry around the AR core.  We therefore concluded that these structures are very common solar features.  Nevertheless, the literature describing these structures is very sparse, providing only a partial view in one or two wavelengths only.

In this work, we presented the first quantitative characterization of the emission properties of the DH associated with active region NOAA~12706, from the chromosphere to the low corona, using IRIS full-Sun mosaics and SDO/AIA images. We complement the results with a preliminary analysis of the SDO/HMI magnetograms.

We found that IRIS scans in the \ion{C}{II} and \ion{Si}{IV} lines of the area around NOAA~12706 show that the DH seen in those lines coincides with the region covered by \ion{Mg}{II} fibrils, although no clear fibrillar pattern could be detected in those TR lines. 

Extending the analysis to SDO/AIA images confirms that DHs are clearly visible in the 171~\r{A} band, as already noted by other authors \citep{wang2011,singh2021}.  However, we also found that the area of the AIA~171~\r{A} DH seems to be significantly larger than in IRIS chromospheric and TR lines, although a direct comparison over the entire region is difficult due to the presence of bright AR loops in the line of sight in many places.  In this work, we therefore distinguish between the DH seen in \ion{Mg}{II} h \& k line core and in AIA~171~\r{A}, which we term ``fibrillar'' and ``coronal'' DH, respectively.  It is not yet clear if and how the two structures are related, but it seems plausible that the ``coronal'' DH encompasses and extends the ``fibrillar'' DH.  This point clearly deserves further investigation, which we are currently carrying out, by analysing for instance a larger set of ARs over a larger range of wavelengths than in the current study. 

In this context, it might be worth noting that the AIA~304~\r{A} line seems to share traits of both kinds of DHs: filtergrams at that wavelength are clearly darker in the fibrillar DH, but the extended area corresponding to the coronal DH is also noticeable darker than the average quiet Sun.

DHs are a clear feature in TR lines. Andretta \& Del Zanna noted their presence around all active regions in TR lines observed by the SOHO/CDS and SOHO/SUMER full-disk scans. Our analysis, in effect, showed that fibrillar DHs are clearly detectable in TR lines, in particular above the Lyman continuum edge; it is therefore unlikely that the appearance of these structures could be due to absorption by cool material, as postulated by \cite{wang2011}. However, the IRIS C II and Si IV lines are formed at temperatures low enough that the coronal DH, observed clearly in the AIA 171 $\AA$ band, is not visible. 

Considering that DHs seen in the AIA~171~\r{A} waveband appear remarkably similar to coronal holes (CHs), we also compared in detail the properties of those two structures, comparing in particular both the ``fibrillar'' and ``coronal'' with the CH seen on the same date at the south pole, using as reference a patch of quiet Sun. Our analysis took into account the effect of average center-to-limb variation which is quite apparent in the intensities of most lines and waveband in the data set.

In addition, we computed average values of the signed and unsigned magnetic field in the regions under studies, using SDO/HMI data taken during the same interval as the IRIS raster scans covering the selected regions.

Our analysis
clearly indicates that the DH and the CH under study are different  structures, as summarised by the following results:
\begin{itemize}
    \item fibrils seen in \ion{Mg}{II} h \& k line core are found only in the area around the associated AR, i.e.\ in the fibrillar DH;
    \item compared to the reference quiet Sun area, the line intensities are different over a wide range of temperatures; this finding is confirmed by the analysis of the emission measures;
    \item the average \ion{C}{II} and \ion{Si}{IV} line profiles in the two region are different; in particular the \ion{Si}{IV} lines exhibit slightly different non-thermal widths compare to both quiet Sun and CH; 
    \item both the fibrillar and coronal DHs in average do not appear to be as markedly unipolar regions as the CH;
    \item the average signed magnetic field in the fibrillar DH, in particular, seem to be vanishingly small, as in the reference quiet Sun patch.
\end{itemize}

We speculate that the above results are valid for all ARs.  If this is the case, some of the differences between DHs and CHs could provide proxies for distinguishing the two structures.  Such proxies could be especially useful in the case of CHs close to ARs, e.g.\ in the case of equatorial CHs. 

In particular, we propose that the different DH and CH emission characteristics in the AIA filters can be used as a simple recipe to distinguish between DHs and CHs.  More specifically, as illustrated by Figs. \ref{fig2c} and \ref{figratios}, DHs are clearly outlined in the AIA~171~\r{A} waveband, while they are nearly indistinguishable from quiescent coronal areas in the AIA~193~\r{A} or AIA~211~\r{A} wavebands.  On the contrary, CHs are dark in all AIA coronal bands, including AIA~193~\r{A} or AIA~211~\r{A}.  

An illustrative example of the above recipe is given by the cases of ARs NOAA 12738 and 12740 shown in Fig. \ref{fig1.0}, which display a $\beta$ and an $\alpha$\footnote{An active region containing a single sunspot or group of sunspots all having the same magnetic polarity.} configuration, respectively. In both cases, fibrils are observed around the brightest part of the AR in the \ion{Mg}{II} h \& h line core. The DH is also clearly detectable around both ARs in the AIA~171 \r{A} images. On the contrary, the AIA~193 \r{A} filtergrams does not show any significantly darker area around the AR core. The CH is instead well visible in both AIA 193 and 171 \r{A} wavebands.

IRIS mosaics are taken only once per month. This means that it is not always possible to check for the presence of fibrils around an AR. However, we highlight that the AIA 304 \r{A} channel, given its peculiarity of showing the annular shadow of fibrils around the AR cores, can be used to have an estimate of the spatial location of the fibrillar DH in the chromosphere when IRIS data are not available. 

DHs are large-scale, long-lasting structures that are clearly associated with active regions. While they apparently persist and perhaps evolve throughout the AR lifetime, they are most likely related to the AR emergence process. There is clearly much less TR plasma connecting the opposite polarities compared to QS regions. The causes could be related to a different magnetic flux emergence in DH which leads to a different coronal heating compared to the quiet Sun areas, or to processes occurring in the corona (or both). 
Regarding the latter, it is worth mentioning that dark regions were found around a few isolated ARs studied by \citet{delzanna2011}. They were, on large scales, relatively unipolar which led the authors to propose an interchange reconnection model between these regions and the hot AR core loops.
It is clear that some interaction must occur between the emerging flux in the cores of ARs and the surrounding pre-existing corona, but it is unclear if the dark halos are related to the interchange reconnection process, especially if it is confirmed that, in terms of magnetic field configurations, they are QS-like regions of largely mixed polarity. 

In conclusion, dark halos remain remarkably common, large, and persistent solar structures, and yet their characteristics and nature is largely unknown.  With this work we have begun to provide a first characterisation of the emission properties of such structures.  
We are carrying out follow-up studies to better determine the properties of  
these features. They include the use of Hinode/EIS (\citealt{culhane2007}) spectroscopic data to 
better characterise the changes in the TR/coronal emission,  
and of Solar Orbiter EUI images (\citealt{rochus2020}), to study DHs at much higher 
spatial resolution than AIA. 
We are also carrying out an analysis of the evolution of DHs in terms of TR/coronal properties and photospheric magnetic field.

}

\begin{acknowledgements}
      This study was partly supported by the Italian agreement ASI-INAF 2021-12-HH.0 “Missione Solar-C EUVST—Supporto scientifico di Fase B/C/D.” 
      GDZ acknowledges financial support from STFC (UK) via the consolidated grants  to the atomic astrophysics group (AAG) at DAMTP, University of Cambridge (ST/T000481/1 and ST/X001059/1). This study has made use of SAO/NASA Astrophysics Data System’s bibliographic services. 
      IRIS is a NASA small-explorer mission developed and operated by LMSAL with mission operations executed at the NASA Ames Research Center and major contributions to downlink communications funded by the European Space Agency and the Norwegian Space Center. CHIANTI is a collaborative project involving George Mason University, the University of Michigan (USA), University of Cambridge (UK) and NASA Goddard Space Flight Center (USA).
\end{acknowledgements}


\newpage
\begin{appendix}

\section{H$\alpha$ GONG images}   
\begin{figure*}
\includegraphics[scale=0.62,trim=20 30 20 15, clip]{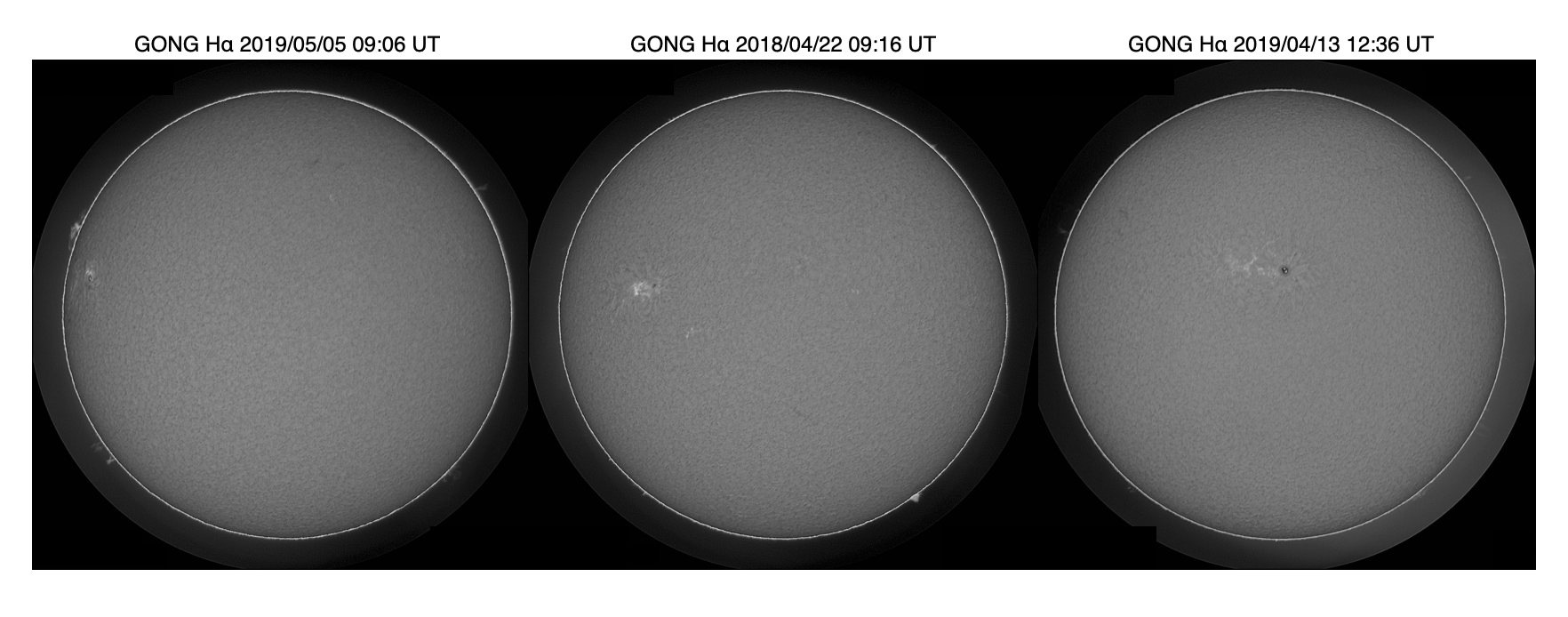}
    \caption{GONG H$\alpha$ images of ARs NOAA 12740 (left panel), NOAA 12706 (middle panel) and NOAA 12738 (right panel). Around the three ARs the fibrils observed in the \ion{Mg}{II} h$_3$ mosaic are glimpsed. No other structures, such as filaments or filament channels, are observed. Data were acquired by the GONG instruments hosted by Learmonth Solar Observatory and Udaipur Solar Observatory, operated by the National Solar Observatory Integrated Synoptic Program (NISP).}
    \label{gong}
\end{figure*}

\section{\ion{Mg}{II} h fibrils}
\begin{figure*}
\centering 
\includegraphics[scale=0.95,trim=0 0 100 675, clip]{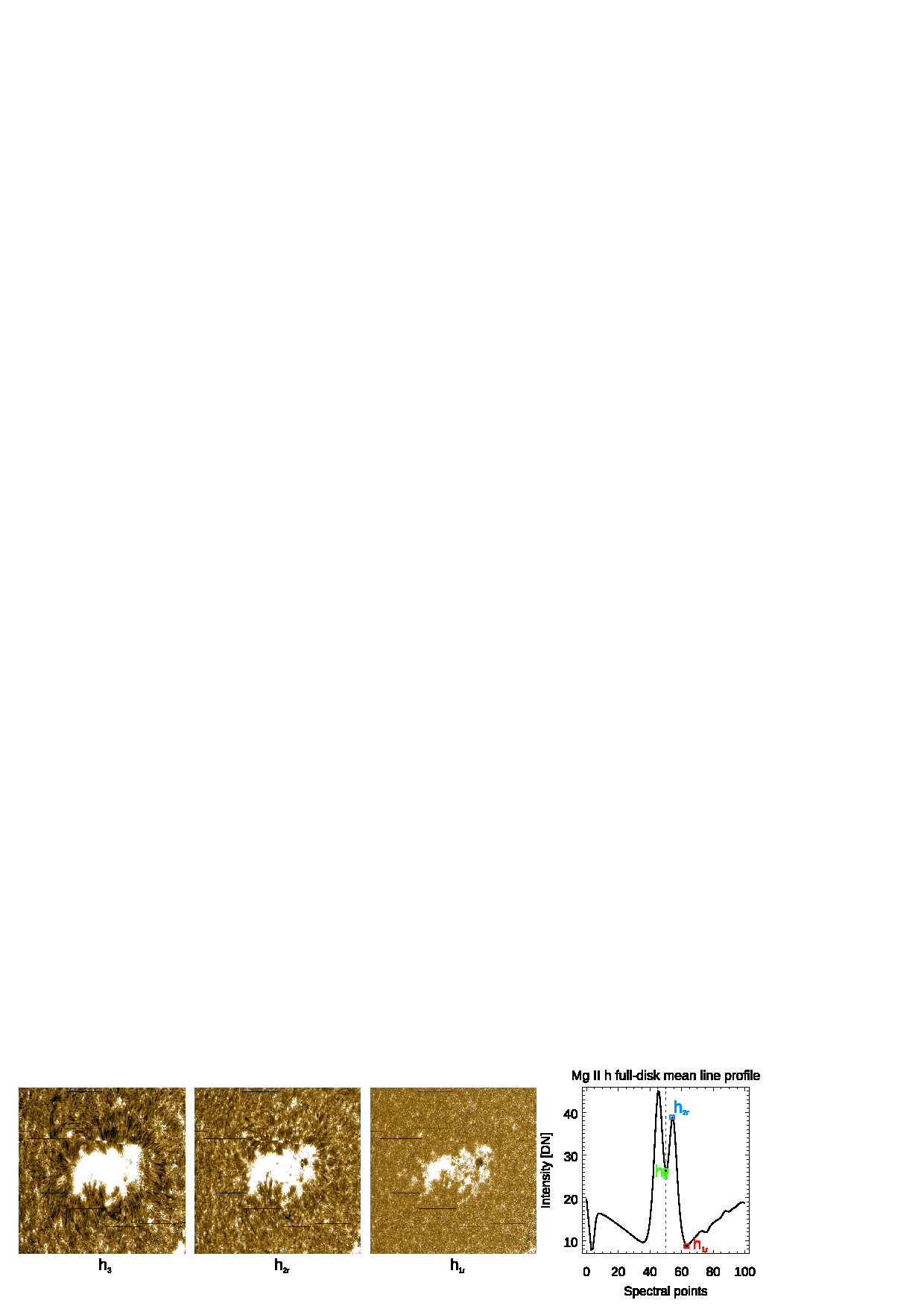}
    \caption{Close-up view of AR NOAA 12706 and its surroundings in the \ion{Mg}{II} h non center-to-limb corrected mosaic at different spectral pixels, i.e., from left to right, at the 50$^{th}$, 54$^{th}$ and 63$^{th}$ spectral point, corresponding respectively to the h$_3$, $h_{2r}$ and $h_{1r}$ in the full-disk mean line profile (right panel). North is up and west is to the right. Field of view has dimensions $500" \times 410"$. It is possible to notice a downward trend in visibility of fibrils as moving away from $h_3$. }
    \label{fibrils}
\end{figure*}

\section{Center-to-limb variations}
\begin{figure*}
    \centering
    \includegraphics[scale=0.45,trim=21 15 8 400,clip]{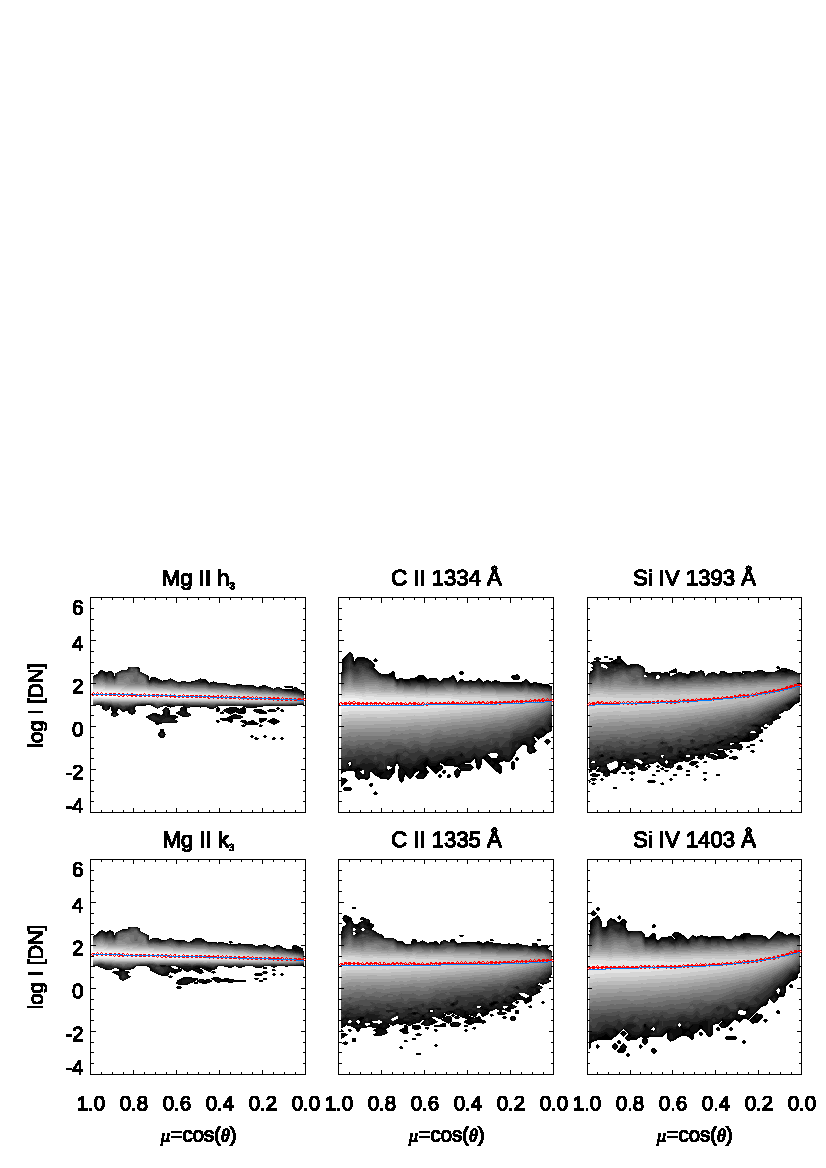}
    \includegraphics[scale=0.45,trim=21 205 8 210,clip]{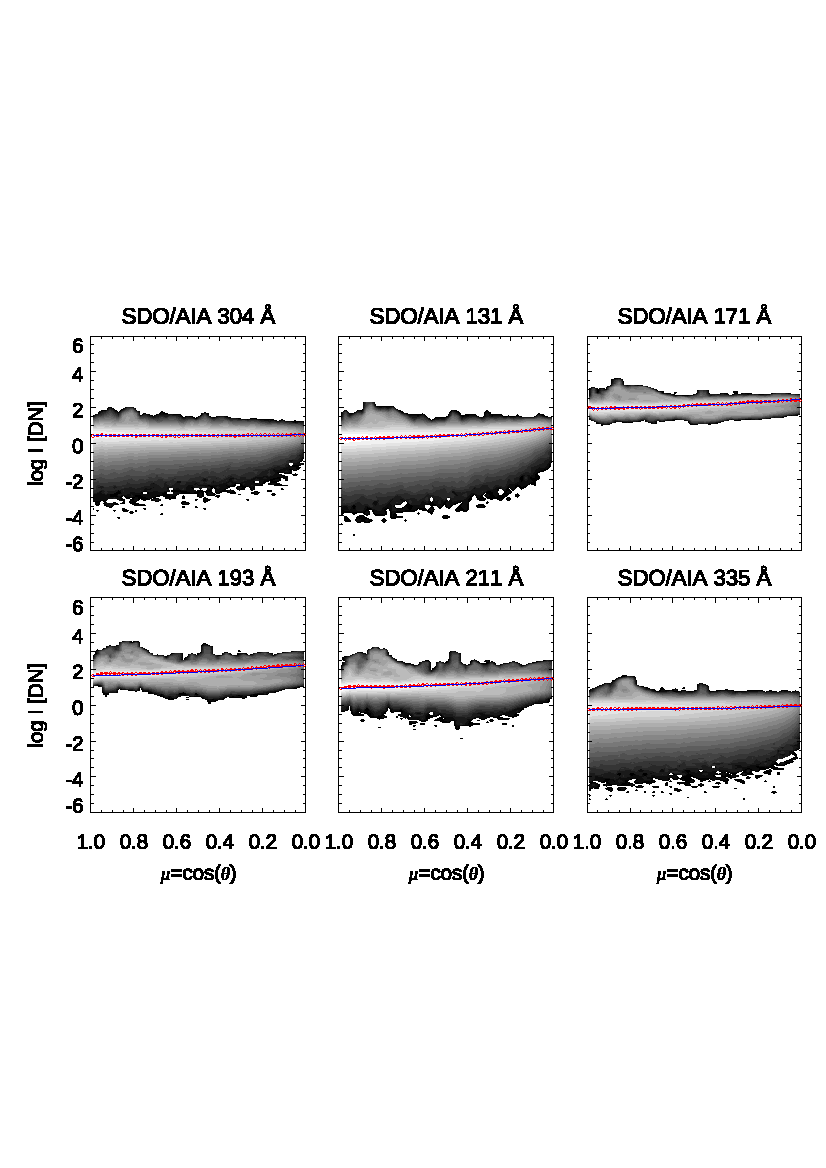}
    \caption{Bi-dimensional histograms vs. $\mu$ are shown for all 2018 April 22 IRIS mosaics (left panels) and AIA filtergrams (right panels) used in the analysis. Each red point is the mode of the histogram in the interval $\Delta\mu=0.02$. The blue curve is the fit obtained from the seven QS dates, as described in Appendix.}
    \label{c2l_figure}
\end{figure*}
In our case, the center-to-limb effects are expected to be relevant in the average intensity measurements only for the CH and the coronal DH, since the QS and the fibrillar DH are at a similar radial distance from the disk center. \\Different center-to-limb effects are evident in all IRIS mosaics and AIA filtergrams. \ion{Mg}{II} h\&k lines are known to present limb darkening (e.g. \citealt{morrill2008}) which has been measured in IRIS mosaics by \citet{gunar2021}. These authors find that the amplitude of the limb darkening decreases as the wavelength range used for the integration. 
As a consequence, we would not expect to find this effect in the h$_3$ or $k_3$ full-disks.
However, we find a non-negligible limb darkening in the h$_3$ and k$_3$ mosaics. All other IRIS mosaics and AIA filtergrams (except AIA 304 \r{A}) show instead more or less obvious center-to-limb brightening, with the \ion{Si}{IV} mosaics and the hotter AIA bands being the most conspicuous.\\
We choose to correct these variations through a standardized empirical method which take into account the center-to-limb trend of each waveband/wavelength. 
The observed intensity I$_{obs}(\mu$ ) for each line/band, as a function of $\mu$, i.e. the cosine of the heliocentric angle $\theta$, is a function of the intensity observed at the disk center I($\mu$ =1 ) hence not modified by line-of-sight effects. This function is expressed with the factor f$_\lambda$($\mu$) which is different for each line or band:
\begin{equation}
    I_{obs,\lambda}(\mu)=f_{\lambda}(\mu)I_{\lambda}(1).
\end{equation}
Therefore, knowing $f_{\lambda}(\mu)$ it is possible to obtain the intensity $I_{\lambda}(\mu=1)$ corrected for the center-to-limb effects.\\
To retrieve the factor $f_{\lambda}(\mu)$ for each IRIS line and AIA waveband we use seven dates of minimum of solar activity (2018-03-26, 2018-09-24, 2018-10-23, 2019-09-22, 2019-10-20, 2020-02-23, 2020-03-23) to characterize the dependence of intensity from radial distance. 
Following an approach similar to that reported in \citet{andretta2014}, we compute the bi-dimensional histogram of intensities as function of $\mu$ in each annulus of size $\Delta\mu=0.02$ starting from the disk center. We exclude from the calculation all pixels with $|Y| < 0.85 \>R_{\odot}$ in order to minimize the contribution of polar CHs that would otherwise affect the quiet Sun profile, above all during the minimum of solar activity. Then we evaluate the peak intensities averaged on the seven dates for each IRIS line and AIA band.  \\
Then, we fit the logarithm of these peak intensities as function of $\mu$ with a cubic function:
\begin{equation}
    log \>I(\mu) = A + B(1-\mu) + C(1-\mu)^2 + D(1-\mu)^3,
\end{equation}
so that the factor $f_{\lambda}(\mu)$ can be simply computed from the following equation:
\begin{equation}
    log \>f_{\lambda}(\mu) = log I(\mu)-log I(1) = B(1-\mu) + C(1-\mu)^2 + D(1-\mu)^3.
\end{equation}
The fits of the seven quiet-Sun day average peak-intensity curves we obtained for IRIS lines and AIA bands are shown in blue in Fig. \ref{c2l_figure} overplotted to the bi-dimensional histograms as function of $\mu$  for the 2018 April 22 date (i.e. AR NOAA 12706 date). Only AIA 94 \r{A} is not shown because its plot is very similar to the AIA 335 \r{A} one. The red lines are the measured positions of intensity peaks of the histograms. 
The cubic function fits very well the observed average center-to-limb variation of intensities.
The seven QS \ion{C}{II} and \ion{Si}{IV} mosaics used to compute $f_{\lambda}(\mu)$ are obtained through line integral. Then, the center-to-limb correction is applied to each wavelength of the 2018 April 22 \ion{C}{II} and \ion{Si}{IV} datacubes under the assumption of optically thin line, which is confirmed by the limb brightening found and shown in Figs. \ref{c2l_figure}.

\end{appendix}

\end{document}